\documentclass[aps,pre,preprint,showpacs,eqsecnum]{revtex4-1}
\usepackage{graphicx}
\usepackage{bm}
\usepackage{amsmath, upgreek}
\begin{document}
\title{Rotational properties of ferromagnetic nanoparticles driven by a precessing magnetic field in a viscous fluid}
\author{T.~V.~Lyutyy$^{1}$}
\email{lyutyy@oeph.sumdu.edu.ua}
\author{S.~I.~Denisov$^{1}$}
\email{denisov@sumdu.edu.ua}
\author{V.~V.~Reva$^{1,2}$}
\author{Yu.~S.~Bystrik$^{1}$}
\affiliation{$^{1}$Sumy State University, 2 Rimsky-Korsakov Street, UA-40007 Sumy, Ukraine\\
$^{2}$NanoBioMedical Center, Adam Mickiewicz University, Umultowska 85, PL-61-614 Pozna\'{n}, Poland}

%\date{submitted to Physical Review B: \today}

\begin{abstract}
We study the deterministic and stochastic rotational dynamics of ferromagnetic nanoparticles in a precessing magnetic field. Our approach is based on the system of effective Langevin equations and on the corresponding Fokker-Planck equation. Two key characteristics of the rotational dynamics, the average angular frequency of precession of nanoparticles and their average magnetization, are of our interest. Using the Langevin and Fokker-Planck equations, we calculate both analytically and numerically these characteristics in the deterministic and stochastic cases, determine their dependence on the model parameters, and analyze in detail the role of thermal fluctuations.
\end{abstract}
\maketitle

\section{INTRODUCTION}
\label{Intr}

Ferromagnetic single-domain nanoparticles possess a number of unique properties, such, for example, as superparamagnetism \cite{Neel, BeLi, Brown}, giant magnetoresistance \cite{Berk, XiJi} and quantum tunnelling of magnetization \cite{ChGu, ThLi, ChTe}. These and other properties provide a basis for numerous current and potential applications of magnetic nanoparticles in data storage \cite{Ross, MTMA, TeTh}, spintronics \cite{BoWe, KKRM}, drug delivery \cite{Ferr, WaAr, MCSS} and hyperthermia \cite{PCJD, ISHK, LFPR, LDHM}, to name only a few. The magnetization dynamics plays a central role in most of these applications, and its characteristics strongly depend on whether the nanoparticles move or not. In the latter case, the time evolution of magnetization can often be described phenomenologically by the Landau-Lifshitz or Landau-Lifshitz-Gilbert equation \cite{LaLi, Gilb}. By adding the thermal torque \cite{Brown} and spin-transfer torque \cite{Slon, Sun} to these equations, they can also be used to study thermal and spin-transfer effects. Within this framework, a wide variety of phenomena, including precessional switching, self-oscillations and thermal relaxation of the nanoparticle magnetization, have already been investigated (for a review, see Ref.~\cite{BMS} and references therein).

Although there is experimental evidence that ferromagnetic nanoparticles can freely rotate even in a solid matrix \cite{BAZP}, it is more obvious that the former case occurs for nanoparticles suspended in a fluid. These systems, also known as ferrofluids, exhibit a number of unique properties \cite{Ros, Oden}. In dilute suspensions, some of these properties are completely determined by the magnetic and mechanical dynamics of independent nanoparticles. Remarkably, in the case of high-anisotropy nanoparticles the magnetization motion describes the nanoparticle rotation as well. Because of its simplicity and efficiency, this approach is very useful for investigating thermal effects in these systems (see, e.g., Refs.~\cite{RaSh, CKW}). In particular, it has been used to predict and study the thermal ratchet effects in ferrofluids subjected to a linearly polarized magnetic field \cite{EMRJ, EnRe}, and to determine the specific absorption rates \cite{RaSt, UsLi}.

In this work, we present a detailed study of the rotational dynamics of highly anisotropic nanoparticles in a precessing magnetic field. We focus on the average angular frequency of nanoparticle rotation and on the average nanoparticle magnetization. The dependencies of these quantities on the model parameters, especially those that exhibit qualitatively different behavior with and without thermal fluctuations, are our main interest here.

The paper is organized as follows. In Sec.~\ref{Mod}, we describe the model and  main approximations and write the basic system of Langevin equations governing the rotational dynamics of ferromagnetic nanoparticles. The corresponding Fokker-Planck equation and the system of effective Langevin equations, which is more convenient than the basic one, are obtained in Sec.~\ref{FPeq}. In the same section, we show that the interpretation of multiplicative Gaussian white noises in effective Langevin equations does not influence the statistical properties of their solution. Section \ref{NoLess} is devoted to the computation of the average angular frequency of precession of nanoparticles and their average magnetization in the deterministic case. The effects of thermal fluctuations are considered in Sec.~\ref{Eff_BR}. Here, we confirm the validity of the system of effective Langevin equations, use it for calculating the above mentioned average characteristics, and discuss the role of thermal fluctuations. Finally, our main conclusions are summarized in Sec.~\ref{Concl}.

\section{MODEL AND BASIC EQUATIONS}
\label{Mod}

We consider a spherical ferromagnetic particle of radius $a$, which rotates in a viscous fluid under a uniform magnetic field $\mathbf{H} = \mathbf{H}(t)$. In our study we use the following assumptions. First, the exchange interaction between magnetic atoms is assumed to be so large that the magnitude $|\mathbf{M}| = M$ of the particle magnetization $\mathbf{M}$ can be considered as a constant parameter. Second, the particle radius is assumed to be so small (less than a few tens of nanometers) that the nonuniform distribution of magnetization becomes energetically unfavorable, i.e., a single-domain state with $\mathbf{M} = \mathbf{M}(t)$ is realized. And third, the magnetic anisotropy field is assumed to be so strong that the magnetization is directed along this field, implying that $\mathbf{M}$ is frozen into the particle body. With these assumptions, the rotational dynamics of a ferromagnetic particle is governed by a pair of coupled equations
\begin{subequations}\label{eqs1}
\begin{eqnarray}
    \dot{\mathbf{M}} &=& \boldsymbol{\upomega}
    \times \mathbf{M},
    \label{eq_a}
    \\[6pt]
    J\dot{\boldsymbol{\upomega}} &=& V\mathbf{M}
    \times \mathbf{H} - 6\eta V \boldsymbol{\upomega}.
    \label{eq_b}
\end{eqnarray}
\end{subequations}
Here, $\boldsymbol{\upomega} = \boldsymbol{\upomega}(t)$ is the angular velocity of the particle, the overdot denotes the time derivative, $J = (2/5)\rho Va^{2}$ is the moment of inertia of the particle, $\rho$ is the particle density, $V = (4/3)\pi a^{3}$ is the particle volume (we associate the hydrodynamic volume of the particle with its own volume), $\eta$ is the dynamic viscosity of the fluid, and the cross denotes the vector product.

The former equation in (\ref{eqs1}) is a special case of the kinematic relation $\dot{\mathbf{a}} = \boldsymbol{\upomega} \times \mathbf{a}$, which holds for an arbitrary frozen vector $\mathbf{a}$ of a fixed length, and the latter one is Newton's second law for rotational motion. The first and second terms in the right-hand side of Eq.~(\ref{eq_b}) are the torques generated by the external magnetic field and viscous fluid at small Reynolds number ($\lesssim 2\times 10^{3}$), respectively. Because the particle size is sufficiently small, the left-hand side of this equation, i.e., the rate of angular momentum $J\boldsymbol{\upomega}$, can safely be neglected in a wide frequency domain. Using this massless approximation and assuming that a random torque $\boldsymbol{\upxi} = \boldsymbol{\upxi}(t)$, which is generated by the thermal motion of fluid molecules, is also applied to a nanoparticle, we obtain
\begin{equation}
    \boldsymbol{\upomega} = \frac{1}{6\eta}\,
    \mathbf{M} \times \mathbf{H} + \frac{1}
    {6\eta V}\, \boldsymbol{\upxi}.
    \label{omega}
\end{equation}
With this result, Eq.~(\ref{eq_a}) reduces to the equation (see, e.g., Ref.~\cite{CKW})
\begin{equation}
    \dot{\mathbf{M}} = - \frac{1}{6\eta }\,
    \mathbf{M} \times (\mathbf{M} \times
    \mathbf{H}) -\frac{1}{6\eta V}\, \mathbf{M}
    \times \boldsymbol{\upxi},
    \label{eq_M}
\end{equation}
which describes the stochastic rotation of ferromagnetic nanoparticles in a viscous fluid. Note that, in spite of the similarity in appearance, Eq.~(\ref{eq_M}) strongly differs from the stochastic Landau-Lifshitz equation describing the magnetization dynamics of fixed nanoparticles. The main difference leading to a qualitatively different behavior of $\mathbf{M}$ is that Eq.~(\ref{eq_M}) does not contain the gyromagnetic term in the deterministic limit. In particular, it is this term that is responsible for the magnetization of nanoparticle systems in a rotating magnetic field \cite{DLH}.

Since the nanoparticle magnetization $M$ does not depend on time, it is convenient to rewrite Eq.~(\ref{eq_M}) in spherical coordinates. To this end, we first represent the magnetization vector as $\mathbf{M} = M\mathbf{m}$ with
\begin{equation}
    \mathbf{m} = \sin{\theta} (\cos{\varphi}\,
    \mathbf{e}_{x} + \sin{\varphi}\, \mathbf{e}_{y})
    + \cos{\theta}\, \mathbf{e}_{z},
    \label{M}
\end{equation}
where $\theta = \theta(t)$ and $\varphi = \varphi(t)$ are the polar and azimuthal angles of the nanoparticle magnetization, respectively, and $\mathbf{e}_{x}$, $\mathbf{e}_{y}$ and $\mathbf{e}_{z}$ are the unit vectors along the corresponding axes of the Cartesian coordinate system $xyz$, whose origin is located at the nanoparticle center. Then, introducing the rescaled random torque as
\begin{equation}
    \boldsymbol{\upzeta} = \left(\frac{1}{12\eta V
    k_{B}T}\right)^{1/2} \boldsymbol{\upxi},
    \label{zeta}
\end{equation}
where $k_{B}$ is the Boltzmann constant and $T$ is the absolute temperature, from Eq.~(\ref{eq_M}) one can obtain the following basic system of stochastic Langevin equations:
\begin{equation}
    \begin{array}{ll}
    \displaystyle \dot{\theta} =\! -\frac{1}{\tau_{1}}
    \frac{\partial w}{\partial \theta}\! - \!\sqrt{
    \frac{2}{\tau_{2}}}\left(\zeta_{x}\sin
    \varphi - \zeta_{y} \cos\varphi\right),
    \\ [10pt]
    \!\displaystyle \dot{\varphi} =\! - \frac{1}{\tau_{1}
    \sin^{2}\!\theta}\frac{\partial w}{\partial
    \varphi}\! - \!\sqrt{\frac{2}{\tau_{2}}}\big[(
    \zeta_{x}\cos \varphi + \zeta_{y}\sin
    \varphi)\cot\theta\! - \!\zeta_{z}\big].
    \end{array}
    \label{eqs2}
\end{equation}
Here, $w = W/M^{2}$, $W=-\mathbf{M} \cdot \mathbf{H}$ is the Zeeman energy density, the dot denotes the scalar product, and
\begin{equation}
    \tau_{1} = \frac{6\eta}{M^{2}}, \quad
    \tau_{2} = \frac{6\eta V}{k_{B}T}
    \label{tau2,3}
\end{equation}
are the time scales characterizing the nanoparticle rotation induced by the external magnetic field and thermal torque, respectively ($\tau_{2}/2$ is also called the Brownian relaxation time). The Cartesian components $\zeta_{\nu}$ ($\nu =x,y,z$) of $\boldsymbol{\upzeta}$ are assumed to be independent Gaussian white noises with zero mean, $\langle \zeta_{\nu} \rangle =0$, and correlation function $\langle \zeta_{\nu}(t) \zeta_{\nu}(t') \rangle = \Delta \delta(t-t')$, where $\langle \cdot \rangle$ denotes averaging over all realizations of Wiener processes $W_{\nu}(t)$ producing noises $\zeta_{\nu}$ (for more details, see the next section), $\Delta$ is the dimensionless noise intensity, and $\delta(t)$ is the Dirac $\delta$ function.

\begin{figure}
    \centering
    \includegraphics[totalheight=8cm]{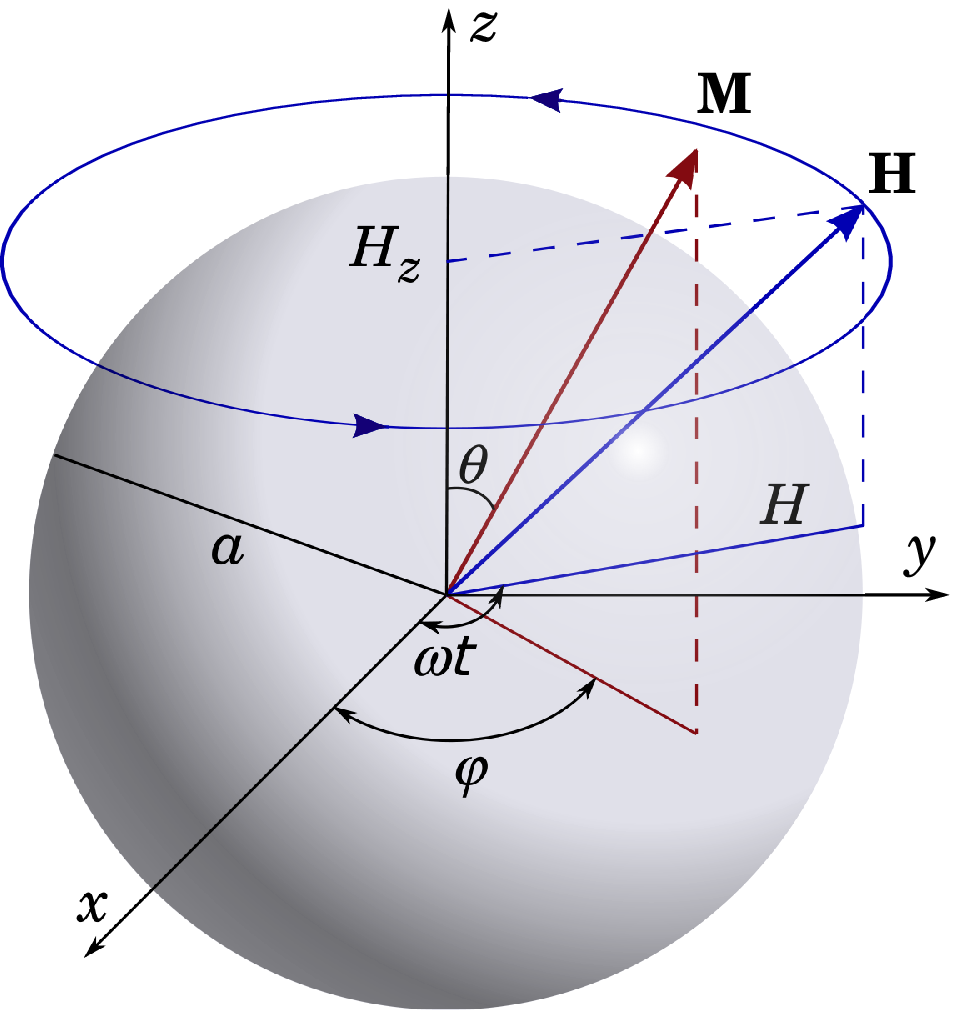}
    \caption{\label{fig1} (Color online) Schematic
    representation of the model. Nanoparticle of
    radius $a$ with frozen magnetization $\mathbf{M}$
    rotates about the origin of the Cartesian coordinate
    system  in a viscous fluid under the action of a
    precessing magnetic field $\mathbf{H}$.}
\end{figure}

Finally, we choose the external magnetic field in the form
\begin{equation}
    \mathbf{H} = H[\cos(\omega t)\mathbf{e}_{x}
    + \sin(\omega t)\mathbf{e}_{y}] + H_{z}
    \mathbf{e}_{z},
    \label{def H}
\end{equation}
where $H$ and $\omega$ are the amplitude and angular frequency of the circularly polarized (rotating) component of $\mathbf{H}$, and $H_{z} \mathbf{e}_{z}$ is the constant component of $\mathbf{H}$ (see Fig.~\ref{fig1}). In this so-called precessing magnetic field, the reduced energy density $w = -\mathbf{m} \cdot \mathbf{h}$ ($\mathbf{h} = \mathbf{H}/M$) is written as
\begin{equation}
    w = - h \sin\theta \cos(\omega t - \varphi)
    - h_{z}\cos\theta
    \label{w}
\end{equation}
with $h = H/M \geq 0$ and $h_{z} = H_{z}/M \geq 0$.

\section{FOKKER-PLANCK EQUATION}
\label{FPeq}

An important feature of the Langevin equations (\ref{eqs2}) is that the noises $\zeta_{\nu}$ are multiplicative, i.e., they are multiplied by functions of the angles $\theta$ and $\varphi$. It is well known (see, e.g., Refs.~\cite{HoLe, Ris}) that the statistical properties of one-dimensional systems described by Langevin equations with multiplicative noises depend on the noises interpretation. In contrast, the statistical properties of some multi-dimensional systems do not depend on how the multiplicative noises are interpreted \cite{MDCH}. Therefore, to determine if the noises interpretation influences the statistical properties of $\theta$ and $\varphi$ and to find the Fokker-Planck equation for the probability density of these angles, Eqs.~(\ref{eqs2}) must be specified more precisely.

For this purpose it is convenient to rewrite the system of stochastic equations (\ref{eqs2}) in the form
\begin{equation}
    \dot{u}_{i} = f_{i}(\mathbf{u}, t) +
    \sum_{j=1}^{3}g_{ij}(\mathbf{u}) \zeta_{j}.
    \label{eq_u}
\end{equation}
Here, $u_{i}$ ($i=1,2$) are the elements of the $2\times 1$ matrix [two-component column vector $\mathbf{u} = \mathbf{u}(t)$] $(u_{i}) = \left(\begin{smallmatrix} u_{1}\\ u_{2} \end{smallmatrix} \right)$ with $u_{1} = \theta$ and $u_{2} = \varphi$, the drift terms $f_{i}( \mathbf{u}, t)$ are the elements of the $2\times 1$ matrix
\begin{equation}
    \left( f_{i} \right) = - \frac{1}{\tau_{1}}
    \begin{pmatrix}
    \partial w/\partial u_{1} \\[6pt]
    \sin^{-2} u_{1}\,\partial w/\partial u_{2}
    \end{pmatrix}
    \label{f_i}
\end{equation}
with $w$ taken from equation (\ref{w}) in which the angles $\theta$ and $\varphi$ are replaced by the variables $u_{1}$ and $u_{2}$, respectively, $\zeta_{1} = \zeta_{x}(t)$, $\zeta_{2} = \zeta_{y}(t)$, $\zeta_{3} = \zeta_{z}(t)$, and the functions $g_{ij}(\mathbf{u})$ are the elements of the $2\times 3$ matrix
\begin{equation}
    \left( g_{ij} \right) = - \sqrt{\frac{2}{\tau_{2}}}
    \begin{pmatrix}
    \sin u_{2} & -\cos u_{2} & 0 \\[6pt]
    \cot u_{1} \cos u_{2} & \cot u_{1}\sin u_{2} & -1
    \end{pmatrix}.\,
    \label{g_ij}
\end{equation}

Then, to specify Eqs.~(\ref{eq_u}), we first assume that the increments $\delta u_{i} = u_{i}(t+\tau) - u_{i}(t)$ of the variables $u_{i}$ at $\tau \ll \min \{\tau_{1}, \tau_{2}\}$ are given by
\begin{equation}
    \delta u_{i} = f_{i}(\mathbf{u}, t)\tau +
    \sum_{j=1}^{3}g_{ij}[\mathbf{u}(t +
    \lambda_{j} \tau)] \delta W_{j},
    \label{incr1}
\end{equation}
where $\lambda_{j} \in [0,1]$ are the parameters characterizing the action of white noises $\zeta_{j}$, and $\delta W_{j} = W_{j}(t+\tau) - W_{j}(t)$ are the increments of Wiener processes $W_{j}(t)$ generating $\zeta_{j}$. Because these noises are assumed to be independent and statistically equivalent, the increments $\delta W_{j}$ can be completely characterized by two conditions
\begin{equation}
    \langle \delta W_{j} \rangle = 0, \quad
    \langle \delta W_{j} \delta W_{l} \rangle
    = \Delta \delta_{jl} \tau
    \label{means}
\end{equation}
with $\delta_{jl}$ being the Kronecker delta. Finally, taking into account that $u_{k}(t + \lambda_{j}\tau) \approx u_{k}(t) + \lambda_{j} \delta u_{k}$ and expanding the last term in Eq.~(\ref{incr1}) to linear order in $\tau$, we obtain
\begin{eqnarray}
    \delta u_{i} &=& f_{i}(\mathbf{u}, t)
    \tau + \sum_{j=1}^{3}g_{ij}(\mathbf{u})
    \delta W_{j} \nonumber \\
    && + \sum_{k=1}^{2}\sum_{j,l=1}^{3}
    \lambda_{j} \frac{\partial g_{ij}
    (\mathbf{u})}{\partial u_{k}}g_{kl}
    (\mathbf{u})\delta W_{l}\delta W_{j}.
    \quad
    \label{incr2}
\end{eqnarray}

Thus, the stochastic equations (\ref{eq_u}) are specified by the difference scheme (\ref{incr2}) in which the noises action is accounted for not only through the increments $\delta W_{j}$ of Wiener processes generating $\zeta_{j}$, but also by the parameters $\lambda_{j}$ realizing an addition connection of the system with these white noises. Since the last term in the right-hand side of Eq.~(\ref{incr2}) is of the order of $\tau$ [cf.\ Eq.~(\ref{means})], this connection is able to strongly modify the statistical characteristics of $u_{i}$. Although the cases with $\lambda_{j} = 0, 1/2,$ and 1 that correspond to the It\^{o} \cite{Ito}, Stratonovich \cite{Strat}, and Klimontovich\cite{Klim} interpretations of Langevin equations, respectively, are usually considered, any other values of $\lambda_{j}$ are allowed from a mathematical point of view. Therefore, the choice of the parameters $\lambda_{j}$ for Eqs.~(\ref{eqs2}) can only be made on physical grounds (see below).

Now, using Eqs.~(\ref{means}) and (\ref{incr2}) and the two-stage procedure of averaging \cite{DVH}, we can derive the Fokker-Planck equation that corresponds to the Langevin equations (\ref{eq_u}). Introducing the probability density $P = P(\mathbf{U}, t)$ that $\mathbf{u}(t) = \mathbf{U}$ as $P = \langle \delta \left( \mathbf{u}(t) - \mathbf{U} \right) \rangle$, where $\mathbf{U}$ is a constant column vector with components $U_{1}$ and $U_{2}$, the straightforward calculations \cite{MDCH} lead to the following Fokker-Planck equation:
\begin{eqnarray}
    \frac{\partial}{\partial t}P &+& \sum_{i=1}^{2}
    \frac{\partial}{\partial U_{i}} \Big(f_{i}
    (\mathbf{U}, t) + \widetilde{f}_{i}(\mathbf{U})\Big)P
    \nonumber \\
    &-& \frac{\Delta}{2}\sum_{i,k=1}^{2}\sum_{j=1}^{3}
    \frac{\partial^{2}}{\partial U_{i}\partial U_{k}}
    g_{ij}(\mathbf{U})g_{kj}(\mathbf{U})P = 0,\quad
    \quad
    \label{FP1}
\end{eqnarray}
where
\begin{equation}
    \widetilde{f}_{i}(\mathbf{U}) = \Delta \sum_{k=1}^{2}
    \sum_{j=1}^{3}\lambda_{j} \frac{\partial g_{ij}
    (\mathbf{U})}{\partial U_{k}} g_{kj}(\mathbf{U})
    \label{def_f}
\end{equation}
are the additional noise-induced drift terms that depend on the interpretation (i.e., values of the parameters $\lambda_{j}$) of stochastic equations (\ref{eq_u}). It should be noted, however, that since the noise $\zeta_{3}$ is additive and so $\partial g_{i3} (\mathbf{U})/ \partial U_{k} = 0$, these terms and, as a consequence, the probability density $P$ do not depend on $\lambda_{3}$.

If the reduced magnetic energy $w$ does not depend on time, then $f_{i}(\mathbf{U}, t) = f_{i}(\mathbf{U})$ and $P$ tends to the equilibrium probability density $P_{0} = P_{0} (\mathbf{U})$ as $t \to \infty$. In this limit, Eq.~(\ref{FP1}) for $P_{0}$ reads
\begin{eqnarray}
    &&\sum_{i=1}^{2} \frac{\partial}{\partial U_{i}}
    \Big(f_{i} (\mathbf{U}) + \widetilde{f}_{i}
    (\mathbf{U}) \Big)P_{0}
    \nonumber \\
    && - \,\frac{\Delta}{2}\sum_{i,k=1}^{2}\sum_{j=1}^{3}
    \frac{\partial^{2}}{\partial U_{i}\partial U_{k}}
    g_{ij}(\mathbf{U})g_{kj}(\mathbf{U})
    P_{0} = 0.\quad
    \label{FPst1}
\end{eqnarray}
It is natural to assume that the solution of this equation is the Boltzmann probability density, which for $h=0$ can be written in the well-known form
\begin{equation}
    P_{0} = \frac{1}{4\pi}\frac{\kappa h_{z}}
    {\sinh (\kappa h_{z})}\sin U_{1}\,
    e^{\kappa h_{z} \cos U_{1}},
    \label{P_st}
\end{equation}
where $\kappa = \tau_{2} /\tau_{1} = M^{2}V/(k_{\mathrm{B}}T)$. Substituting Eq.~(\ref{P_st}) into Eq.~(\ref{FPst1}) and using the definitions (\ref{f_i}) and (\ref{g_ij}), we straightforwardly obtain
\begin{eqnarray}
    \lambda_{1} &+& \lambda_{2} -1 + \frac{\kappa h_{z}}
    {\Delta}[2 + \Delta(\lambda_{1} + \lambda_{2}
    - 3)] \cos U_{1}
    \nonumber \\ [3pt]
    &-& (\lambda_{2} - \lambda_{1})\cos (2U_{2})
    \bigg( \frac{4}{\sin^{2}U_{1}} + \kappa h_{z} \cos
    U_{1} - 1 \bigg)
    \nonumber \\ [3pt]
    &+& \frac{(\kappa h_{z})^{2}}{\Delta}(\Delta - 1)
    \sin^{2} U_{1} = 0.
    \label{cond1}
\end{eqnarray}
This condition holds for all possible values of the variables $U_{1}$ and $U_{2}$ ($0\leq U_{1} \leq \pi,\, 0\leq U_{2} < 2\pi$) and parameter $\kappa h_{z}$ ($0 \leq \kappa h_{z}< \infty$), i.e., Eq.~(\ref{P_st}) is the solution of the Fokker-Planck equation (\ref{FPst1}), only if
\begin{equation}
    \Delta =1, \quad \lambda_{1} = \lambda_{2} =
    \frac{1}{2}.
    \label{cond2}
\end{equation}
Thus, if Eqs.~(\ref{eqs2}) with Gaussian white noises of unit intensity are interpreted in the Stratonovich sense, the random rotations of nanoparticles are characterized by Boltzmann statistics at long times.

Now, using the conditions (\ref{cond2}) and introducing the variables $\Theta = U_{1}$ and $\Phi = U_{2}$, the Fokker-Planck equation (\ref{FP1}) can be rewritten in the form
\begin{eqnarray}
    \frac{\partial P}{\partial t} &-& \frac{1}{\tau_{1}}
    \frac{\partial}{\partial \Theta}\bigg( \frac{\partial
    w}{\partial \Theta} - \frac{1}{\kappa}
    \cot \Theta\bigg)P - \frac{1}{\tau_{1} \sin^{2}\Theta}
    \frac{\partial}{\partial \Phi} \bigg( \frac{\partial w}
    {\partial \Phi}P \bigg)
    \nonumber \\[6pt]
    &-&  \frac{1}{\tau_{2}}\bigg( \frac{\partial^{2}}
    {\partial \Theta^{2}} + \frac{1}{\sin^{2}\Theta}
    \frac{\partial^{2}}{\partial \Phi^{2}}\bigg)P =0.
    \quad
    \label{FP2}
\end{eqnarray}
We assume that the solution $P = P(\Theta, \Phi, t)$ of this equation is properly normalized, i.e.,
\begin{equation}
    \int_{0}^{\pi} d\Theta \int_{0}^{2\pi}
    d\Phi\, P(\Theta, \Phi, t) = 1,
    \label{cond3}
\end{equation}
and satisfies the initial condition $P(\Theta, \Psi, 0) = \delta (\Theta - \Theta_{0}) \delta (\Phi - \Phi_{0})$ with $\Theta_{0} = \theta (0)$ and $\Phi_{0} = \varphi(0)$.

\subsection{Effective Langevin equations}
\label{Lang}

According to the above results, the basic Langevin equations (\ref{eqs2}) should be interpreted in the Stratonovich sense. Due to this fact and because the system of two equations (\ref{eqs2}) contains three Gaussian white noises, the study of the rotational dynamics of nanoparticles by the numerical solution of these equations is not quite practical. Therefore, it is convenient to use, instead of Eqs.~(\ref{eqs2}), a system of effective Langevin equations satisfying the following requirements. First, the statistical properties of solutions of the basic and effective equations must be the same and, second, the effective equations must be interpreted in the Ito sense and contain two rather than three independent  Gaussian white noises. It has been shown \cite{RaEn} that the corresponding system of effective Langevin equations can be written as
\begin{equation}
    \begin{array}{ll}
    &\dot{\theta} = \displaystyle -\frac{1}{\tau_{1}}
    \frac{\partial w}{\partial \theta} + \frac{1}
    {\tau_{2}}\cot \theta + \sqrt{\frac{2}{\tau_{2}}}
    \,\mu_{1},
    \\ [14pt]
    &\!\dot{\varphi} = - \displaystyle \frac{1}{\tau_{1}
    \sin^{2}\theta}\frac{\partial w}{\partial
    \varphi} + \sqrt{\frac{2}{\tau_{2}}} \frac{1}
    {\sin\theta}\, \mu_{2},
    \end{array}
    \label{eff_Lan1}
\end{equation}
where $\mu_{i} = \mu_{i}(t)$ ($i = 1,2$) are independent Gaussian white noises with zero means, $\langle \mu_{i}(t) \rangle =0$, and delta correlation functions,  $\langle \mu_{i}(t)\mu_{i}(t') \rangle  = \delta(t-t')$. Note that the similar system of effective Langevin equations, which corresponds to the Landau-Lifshitz-Gilbert equation describing the stochastic dynamics of magnetization in single-domain ferromagnetic nanoparticles embedded into a solid matrix, has been proposed in Ref.~\cite{DSTH}.

According to the results of Ref.~\cite{RaEn}, the probability density of the solution of Eqs.~(\ref{eff_Lan1}) satisfies the Fokker-Planck equation (\ref{FP2}). As a consequence, the rotational properties of ferromagnetic nanoparticles can be described either by Eqs.~(\ref{eqs2}) interpreted in the Stratonovich sense or, equivalently, by Eqs.~(\ref{eff_Lan1}) interpreted in the Ito sense. A remarkable feature of the latter equations is that, independent of their interpretation, the corresponding Fokker-Planck equation is given by Eq.~(\ref{FP2}). Indeed, rewriting Eqs.~(\ref{eff_Lan1}) in the form
\begin{equation}
    \dot{u}_{i} = F_{i}(\mathbf{u}, t) +
    \sum_{j=1}^{2}G_{ij}(\mathbf{u}) \mu_{j},
    \label{eq_u2}
\end{equation}
where
\begin{equation}
    \left( F_{i} \right) = - \frac{1}{\tau_{1}}
    \begin{pmatrix}
    \partial w/\partial u_{1} - (1/\kappa)\cot u_{1}
    \\[6pt]
    \sin^{-2} u_{1}\,\partial w/\partial u_{2}
    \end{pmatrix}
    \label{F_i}
\end{equation}
and
\begin{equation}
    \left( G_{ij} \right) = \sqrt{\frac{2}{\tau_{2}}}
    \begin{pmatrix}
    1 &  0 \\[6pt]
    0 & 1/\sin u_{1}
    \end{pmatrix},
    \label{G_ij}
\end{equation}
one can straightforwardly verify that the condition
\begin{equation}
    \sum_{k=1}^{2}\frac{\partial G_{ij}(\mathbf{u})}
    {\partial u_{k}} G_{kj}(\mathbf{u}) = 0
    \label{cond4}
\end{equation}
holds for all  $i$ and $j$. This means \cite{MDCH} [see also Eqs.~(\ref{FP1}) and (\ref{def_f})] that the Fokker-Planck equation associated with Eqs.~(\ref{eff_Lan1}), i.e., Eq.~(\ref{FP2}), does not depend on the parameters $\lambda_{j}$  providing a quantitative interpretation of stochastic equations (\ref{eff_Lan1}). While this conclusion is obvious for $\lambda_{1}$ (because the noise $\mu_{1}$ is additive), the independence of Eq.~(\ref{FP2}) on $\lambda_{2}$ is rather surprising (because the noise $\mu_{2}$ is multiplicative). We note in this context that, in contrast to the univariate case, there always exists a class of multivariate Langevin equations with multiplicative Gaussian white noises whose interpretation does not influence the corresponding Fokker-Planck equations \cite{MDCH}. The above results show that Eqs.~(\ref{eff_Lan1}) belong to this unique class of Langevin equations.

\section{NOISELESS CASE}
\label{NoLess}

Before proceeding with the study of thermal effects, we first briefly discuss the deterministic (noiseless) case.  In this case, taking the limit $\tau_{2} \to \infty$ and using Eq.~(\ref{w}), from Eqs.~(\ref{eqs2}) we obtain the following system of deterministic equations:
\begin{equation}
    \begin{array}{ll}
    \displaystyle \,\tau_{1}\dot{\theta} =
    h\cos\theta \cos\psi -h_{z} \sin\theta,
    \\ [10pt]
    \displaystyle \tau_{1}\dot{\psi} = \Omega -
    h\,\frac{\sin\psi}{\sin\theta},
    \end{array}
    \label{det_eqs}
\end{equation}
where $\psi = \omega t - \varphi$ is the lag angle and $\Omega = \omega \tau_{1}$ is the dimensionless frequency of the rotating component of the magnetic field (\ref{def H}). Assuming that $\theta \to \theta_{\infty} = \mathrm{const}$ and $\psi \to \psi_{\infty} = \mathrm{const}$ as $t \to \infty$, the above differential equations in the long-time limit are reduced to a set of trigonometric equations
\begin{equation}
    \begin{array}{ll}
    h\cos\theta_{\infty} \cos\psi_{\infty}
    - h_{z} \sin\theta_{\infty} = 0,
    \\ [12pt]
    \Omega \sin \theta_{\infty}  - h\sin
    \psi_{\infty} = 0.
    \end{array}
    \label{stat_eqs}
\end{equation}
Because the solutions of these equations in the cases with $h_{z} > 0$ and $h_{z} = 0$ can be quite different, we consider them separately.

\subsection{$\bm{h_{z} > 0}$}

Using Eqs.~(\ref{stat_eqs}) and the condition $h_{z} > 0$, it can easily be shown that the stationary solution of the deterministic equations (\ref{det_eqs}) is given by
\begin{equation}
    \begin{pmatrix}
    \theta_{\infty}
    \\[6pt]
    \psi_{\infty}
    \end{pmatrix}\!
    = \arcsin\! \left[\frac{1}{\sqrt{2}}\sqrt{\Gamma^{2} -
    \sqrt{\Gamma^{4} - 4\Omega^{2}h^{2}}}
    \begin{pmatrix}
    1/\Omega
    \\[6pt]
    1/h
    \end{pmatrix}\!\right]
    \label{psi_theta}
\end{equation}
($\Gamma^{2} = \Omega^{2} + h^{2} + h_{z}^{2}$). It can also be proven that this solution is stable with respect to small perturbations of angles $\theta$ and $\psi$. Therefore, the solution of Eqs.~(\ref{det_eqs}) at $h_{z} > 0$ always tends to the stationary solution (\ref{psi_theta}) as $t \to \infty$. In particular, the $z$-component of the reduced nanoparticle magnetization in the long-time limit, $m_{z} = \cos \theta_{ \infty}$, can always be represented in the form
\begin{equation}
    m_{z} = \frac{1} {\sqrt{2}\Omega}
    \sqrt{2\Omega^{2} - \Gamma^{2} +
    \sqrt{\Gamma^{4} - 4\Omega^{2}h^{2}}}.
    \label{m_z}
\end{equation}

\begin{figure}
    \centering
    \includegraphics[height=5cm,width=8.5cm]{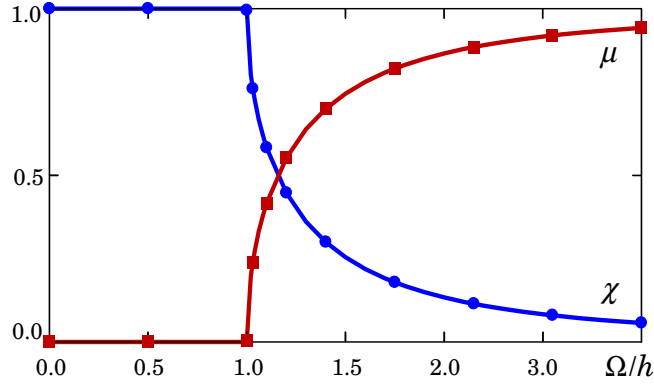}
    \caption{\label{fig2} (Color online) Nanoparticle
    magnetization $\mu$ and nanoparticle angular frequency
    $\chi$ as functions of the reduced magnetic field
    frequency $\Omega/h$. The solid curves represent the
    analytical results (\ref{mu}) and (\ref{chi}), and the
    symbols indicate the numerical results obtained at $t=
    10^{7}\tau_{1}$ by solving Eqs.~(\ref{det_eqs}) for
    maghemite nanoparticles in water. The numerical values
    of $\mu$ and $\chi$, which are calculated at $h_{z} =
    10^{-3}$ and $h_{z}= 0$, respectively, do not depend on
    the initial values $\theta(0)$ and $\psi(0)$ of the polar
    and lag angles.}
\end{figure}

In general, $m_{z}$ as a function of the parameters $\Omega$, $h$ and $h_{z}$ exhibits the expected limiting behavior: $m_{z} \to h_{z}/ \sqrt{h^{2} + h_{z}^{2}}$ as $\Omega \to 0$, $m_{z} \to 1$ as $\Omega \to \infty$, $h \to 0$ or $h_{z} \to \infty$, and $m_{z} \to 0$ as $h \to \infty$. But the dependence of $m_{z}$ on $\Omega$ and $h$ under the condition that $h_{z} \to 0$ is not so obvious. Indeed, from Eq.~(\ref{m_z}) one obtains
\begin{equation}
    \mu = \lim_{h_{z} \to 0} m_{z} = \left\{\!\!
    \begin{array}{ll}
    0, & \Omega/h < 1
    \\ [6pt]
    \sqrt{1 - (h/\Omega)^{2}},
    & \Omega/h \geq 1,
    \end{array}
    \right.
\label{mu}
\end{equation}
i.e., the nanoparticle magnetization $\mu$ depends only on the ratio $\Omega/h$ and, what is more important, the behavior of $\mu$ in the regions $\Omega/h < 1$ and $\Omega/h \geq 1$ is qualitative different. As illustrated in Fig.~\ref{fig2}, the numerical solution of Eqs.~(\ref{det_eqs}) obtained by the fourth-order Runge-Kutta method confirms this theoretical result.
Here and in the following, the numerical calculations are carried out for maghemite ($\gamma$-$ \mathrm{Fe}_{2} \mathrm{O}_{3}$) nanoparticles in water. In this case, the saturation magnetization of nanoparticles, dynamic viscosity of water and characteristic time $\tau_{1}$ at room temperature $T = 298\, \mathrm{K}$ are given by $4\pi M = 3.89\times 10^{3}\, \mathrm{G}$, $\eta = 8.90 \times 10^{-3}\, \mathrm{P}$ and $\tau_{1} = 5.56 \times 10^{-7} \, \mathrm{s}$, respectively.

\subsection{$\bm{h_{z} = 0}$}

Equation (\ref{mu}) shows that the case with $h_{z}=0$ is special. If $\Omega/h <1$ then Eqs.~(\ref{stat_eqs}) or Eq.~(\ref{psi_theta}) yield
\begin{equation}
    \theta_{\infty} = \frac{\pi}{2}, \quad
    \psi_{\infty} = \arcsin\! \left( \frac{
    \Omega}{h}\right),
    \label{psi_theta2}
\end{equation}
and, as in the previous case, this solution is stable. In contrast, if $\Omega/h \geq 1$ then the steady-state solution of Eqs.~(\ref{det_eqs}) is periodic in time with period $t_{\mathrm{st}} = \pi \tau_{1}/\sqrt{\Omega^{2} - h^{2}}$ (periodic regime of rotation) \cite{HaBu}. More precisely, in this case the angles $\theta(t)$ and $\psi(t)$ are changed in such a way that $\theta(t + t_{\mathrm{st}}) = \theta(t)$ and $\psi(t + t_{\mathrm{st}}) = \pi + \psi(t)$. Using these results, it is possible to determine the reduced angular frequency of nanoparticles, which is defined as $\chi = (1/\omega) \lim_{t \to \infty} \varphi(t)/t$. Indeed, since $\varphi(t) = \omega t - \psi(t)$, from this definition one gets $\chi = 1$ for $\Omega/h <1$ and $\chi = 1 - \pi/(\omega t_{\mathrm{st}})$ for $\Omega/h \geq 1$, i.e.,
\begin{equation}
    \chi = \left\{\!\! \begin{array}{ll}
    1, & \Omega/h < 1
    \\ [6pt]
    1 - \sqrt{1 - (h/\Omega)^{2}},
    & \Omega/h \geq 1.
    \end{array}
    \right.
    \label{chi}
\end{equation}
This dependence of $\chi$ on $\Omega/h$ is also in excellent agreement with the numerical results, as shown in Fig.~\ref{fig2}.

\begin{figure}
    \centering
    \includegraphics[height=5cm,width=8.5cm]{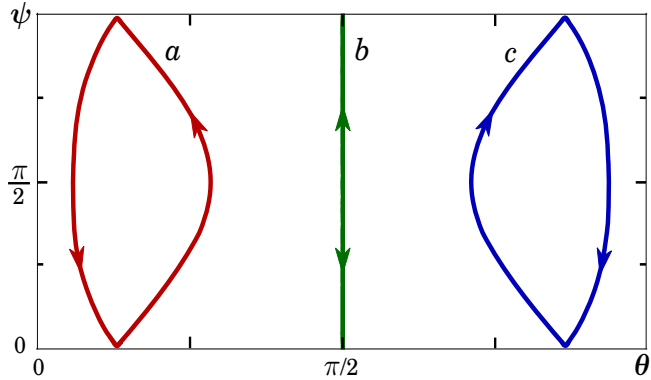}
    \caption{\label{fig3} (Color online) Examples of
    steady-state trajectories of the rotational motion
    of nanoparticles illustrating their dependence on
    the initial polar angle $\theta(0)$. The trajectories
    are obtained from the numerical solution of
    Eqs.~(\ref{det_eqs}) for $\Omega= 2.5 $, $h= 1$,
    $h_{z}=0$, $\psi(0) = 0$, and $\theta(0) = 0.4\;
    (a)$, $\theta(0) = \pi/2\; (b)$,  $\theta(0) =  2.7\;
    (c)$. In this figure, the angles are measured in radians,
    the lag angle $\psi(t)$ is reduced to the interval
    $[0,\pi]$, the arrows indicate the direction of
    time evolution, and the degenerate trajectory $b$
    corresponds to the nanoparticle rotation in the
    $xy$-plane [i.e., $\theta(t) = \pi/2$].}
\end{figure}

Comparing Eq.~(\ref{chi}) with Eq.~(\ref{mu}), we can see that the nanoparticle magnetization $\mu$ and the nanoparticle angular frequency $\chi$ are connected in a remarkably simple way: $\mu + \chi = 1$. It should be noted that, although the steady-state dynamics of the unit magnetization vector $\mathbf{m}$ at $h_{z}=0$ and $\Omega/h \geq 1$ may strongly depend on the initial direction of this vector (see Fig.~\ref{fig3} for illustration), there is no initial-state dependence for $\chi$. Thus, the condition $\mu + \chi = 1$ is universal in the sense that it holds for all possible values of the reduced magnetic field frequency $\Omega/h$ and does not depend on the initial values $\theta(0)$ and $\psi(0)$ of the polar and lag angles.

To avoid confusion in interpreting the above condition, we first recall that at $h_{z} > 0$ the stationary solution (\ref{psi_theta}) of Eqs.~(\ref{det_eqs}) is stable for all values of the ratio $\Omega/h$. In this case, $m_{z}$ is given by Eq.~(\ref{m_z}), $\chi =1$ and, in addition, $m_{z}$ can be approximated by Eq.~(\ref{mu}) if $h_{z}$ is small but nonzero. In contrast, since at $h_{z}=0$ the stationary solution (\ref{psi_theta2}) of Eqs.~(\ref{det_eqs}) is stable only if $\Omega/h<1$ and these equations at $\Omega/h \geq 1$ have a periodic steady-state solution, in this case the nanoparticle angular frequency is given by Eq.~(\ref{chi}) and the nanoparticle magnetization equals zero: $\mu=0$. The last result following from the definition $\mu = (1/t_{ \mathrm{st}}) \int_{0}^{t_{ \mathrm{st}}} dt \cos \theta(t)$, which accounts for the existence of periodic solution of Eqs.~(\ref{det_eqs}) at $\Omega/h \geq 1$, shows that the rotating magnetic field (when $h_{z}=0$) does not magnetize the reference systems. Thus, the condition $\mu + \chi = 1$ holds if $\mu$ is associated with $m_{z}$ at $h_{z} \ll 1$ (not at $h_{z}=0$) and $\chi$ is taken at $h_{z} = 0$. We note in this context that the rotational regime of nanoparticles, which exists at infinitesimally small $h_{z}$, is completely destroyed by thermal fluctuations (see below).

\section{EFFECTS OF THERMAL FLUCTUATIONS}
\label{Eff_BR}

Next, to study thermal effects in the rotational dynamics of ferromagnetic nanoparticles with frozen magnetization, we solve analytically the Fokker-Planck equation (\ref{FP2}) and numerically the system of effective Langevin equations (\ref{eff_Lan1}).

\subsection{Steady-state solution of the Fokker-Planck equation}

The results obtained for the noiseless case suggest that, depending on the model parameters, the steady-state solution $P_{\mathrm{st}}$ of the Fokker-Planck equation (\ref{FP2}) at $t \to \infty$ can be represented as a function of two variables $\Theta$ and $\Psi = \omega t - \Phi$, i.e., $P_{\mathrm{st}} =P_{\mathrm{st}} (\Theta, \Psi)$. Using the relation $\partial P_{\mathrm{st}} /\partial t = \omega \partial P_{\mathrm{st}}/\partial \Psi$ and Eq.~(\ref{FP2}), we can find the equation for the steady-state probability density $P_{\mathrm{st}}$ directly from Eq.~(\ref{FP2}). For brevity, it is convenient to write this equation in the operator form
\begin{equation}
    \hat{L}P_{\mathrm{st}} = \kappa \Omega\frac{
    \partial P_{\mathrm{st}}} {\partial \Psi},
    \label{FPst2}
\end{equation}
where the Fokker-Planck operator $\hat{L}$ is defined as
\begin{eqnarray}
    \hat{L}P_{\mathrm{st}} &=& \frac{\partial}{\partial
    \Theta}\bigg(\!\kappa \frac{\partial w}{\partial
    \Theta} - \cot \Theta\! \bigg)P_{\mathrm{st}}
    + \frac{\kappa}{\sin^{2} \Theta} \frac{\partial}
    {\partial \Psi}\bigg( \frac{\partial w}{\partial
    \Psi} P_{\mathrm{st}} \! \bigg)
    \nonumber \\[6pt]
    && + \bigg( \frac{\partial^{2}}{\partial
    \Theta^{2}} + \frac{1}{\sin^{2}\Theta}
    \frac{\partial^{2}}{\partial \Psi^{2}}\bigg)
    P_{\mathrm{st}}
    \label{def_L}
\end{eqnarray}
with $w = w(\Theta, \Psi) = - h \sin\Theta \cos\Psi - h_{z}\cos\Theta$. In particular, if $\Omega = 0$, then $P_{\mathrm{st}}$ is reduced to the equilibrium Boltzmann probability density
\begin{equation}
    P_{0} = P_{0}(\Theta, \Psi) = \frac{1}{Z}
    \sin \Theta \, e^{-\kappa w(\Theta, \Psi)}
    \label{P_01}
\end{equation}
($Z = \int_{0}^{\pi}d\Theta \int_{0}^{2\pi}d\Psi \sin \Theta \, e^{-\kappa w(\Theta, \Psi)}$), which is the normalized solution of the equation $\hat{L}P_{0} = 0$.

Assuming that $\kappa \Omega \ll 1$, the steady-state probability density $P_{\mathrm{st}}$ can be expanded in a power series of $\kappa \Omega$. In the linear approximation in $\kappa \Omega$ this expansion yields
\begin{equation}
    P_{\mathrm{st}} = (1 + \kappa \Omega F)P_{0},
    \label{P_st1}
\end{equation}
where, according to Eq.~(\ref{FPst2}), $F = F(\Theta, \Psi)$ is the solution of the following equation:
\begin{equation}
    \hat{L}(P_{\mathrm{0}}F) = \frac{\partial
    P_{0}}{\partial \Psi}.
    \label{eq_F1}
\end{equation}
Since the probability densities $P_{\mathrm{st}}$ and $P_{0}$ are normalized, the function $F$ must also satisfy the condition
\begin{equation}
    \int_{0}^{\pi}d\Theta \int_{0}^{2\pi}d\Psi
    F(\Theta, \Psi)P_{0}(\Theta, \Psi) =0.
    \label{norm_F}
\end{equation}

In what follows, we restrict ourselves to the case when $h_{z} = 0$ and $\kappa h \ll 1$. Then, using Eqs.~(\ref{P_01}) and (\ref{eq_F1}), it is not difficult to show that in the main approximation in $\kappa h$ the function $F$ is determined by the equation
\begin{equation}
    \frac{1}{\sin \Theta} \frac{\partial}
    {\partial \Theta}\bigg(\! \sin \Theta
    \frac{\partial F}{\partial \Theta}\bigg)
    + \frac{1}{\sin^{2} \Theta} \frac{
    \partial^{2} F}{\partial \Psi^{2}} =
    - \kappa h \sin \Theta \sin \Psi.
    \label{eq_F2}
\end{equation}
The solution of this equation, which vanishes as $\kappa h \to 0$, has the form
\begin{equation}
    F = \frac{1}{2} \kappa h \sin\Theta \sin\Psi.
    \label{F}
\end{equation}
Therefore, taking into account that, up to quadratic order in $\kappa h$, $Z = 4\pi (1 + \kappa^{2}h^{2}/6)$ and
\begin{equation}
    P_{0} = \frac{\sin\Theta}{4\pi}\! \Big[1 + \kappa h
    \sin\Theta \cos\Psi -\frac{\kappa^{2}h^{2}}
    {6}(1 - 3\sin^{2}\Theta \cos^{2}\Psi) \Big]\!,
    \label{P_02b}
\end{equation}
from Eq.~(\ref{P_st1}) one immediately gets
\begin{equation}
    P_{\mathrm{st}} = P_{0} + \frac{1}{8\pi}
    \kappa^{2}h \Omega \sin^{2}\Theta \sin\Psi.
    \label{P_st2}
\end{equation}
To avoid any confusion, we emphasize that this result is obtained under the assumption that $P_{\mathrm{st}} =P_{\mathrm{st}} (\Theta, \Psi)$, $h_{z}=0$ and $\kappa \max{(\Omega, h)} \ll 1$.

\subsection{Simulation results}

Introducing the dimensionless time $\tilde{t} = t/\tau_{1}$ and using Eq.~(\ref{w}), the system of effective Langevin equations (\ref{eff_Lan1}) in the rotating frame can be written as
\begin{equation}
    \begin{array}{ll}
    &\displaystyle \frac{d\theta}{d\tilde{t}} =
    h\cos\theta \cos\psi\! - \!h_{z}\sin\theta +
    \!\frac{1}{\kappa}\! \cot\theta + \!\sqrt{
    \frac{2}{\kappa}} \tilde{\mu}_{1},
    \\ [12pt]
    &\!\displaystyle \frac{d\psi}{d\tilde{t}} =
    \Omega - h\, \frac{\sin\psi}{\sin\theta} -
    \!\sqrt{\frac{2}{\kappa}}\, \frac{1}{\sin\theta}
    \, \tilde{\mu}_{2},
    \end{array}
    \label{eff_Lan2}
\end{equation}
where $\tilde{\mu}_{i} = \tilde{\mu}_{i}(\tilde{t}) = \sqrt{\tau_{1}}\, \mu_{i}( \tilde{t} \tau_{1})$ ($i=1,2$) are dimensionless Gaussian white noises with $\langle \tilde{\mu}_{i}(\tilde{t}) \rangle =0$ and $\langle \tilde{\mu}_{i} (\tilde{t}) \tilde{\mu}_{i} (\tilde{t}') \rangle = \delta( \tilde{t} - \tilde{t}')$. It is this system of equations that we used in our simulations.

In order to verify if thermal fluctuations are properly taken into account in the effective Langevin equations (\ref{eff_Lan2}), we solved these equations by the Runge-Kutta method and calculated the quantity
\begin{equation}
    \sigma = \frac{8\pi}{\kappa^{2} h^{2}}
    (P_{\mathrm{st}} - P_{0})|_{\Psi = \pi/2}.
    \label{sigma}
\end{equation}
According to Eq.~(\ref{P_st2}), this quantity characterizes the difference between the steady-state (when $\Omega \neq 0$) and equilibrium (when $\Omega = 0$) probability densities at $\Psi = \pi/2$ and is expressed as  $\sigma = (\Omega/h) \sin^{2}\Theta$. The numerical results for $\sigma$ as a function of $\Theta$ are obtained by solving Eqs.~(\ref{eff_Lan2}) for $\tau_{1} = 5.56 \times 10^{-7} \,\mathrm{s}$, $\kappa = 8$, $h_{z} =0$ and different values of $h$ and $\Omega$. The solutions of these equations, i.e., the pairs of angles $\theta_{n} = \theta( \tilde{t}_{n})$ and $\psi_{n} = \psi( \tilde{t}_{n})$, are determined at the moments of time $\tilde{t}_{n} = 10^{5} + n \Delta \tilde{t}$ (this choice of the initial time guaranties that the transient processes are completed) with $n = \overline{1,N}$, $N = 10^{11}$ and $\Delta \tilde{t} = 10^{-2}$. Finally, the numerical values of the probability densities $P_{\mathrm{st}} |_{\Psi = \pi/2}$ and $P_{0}|_{\Psi = \pi/2}$ are calculated as $N_{m}|_{\Omega \neq 0}/(\Delta\Theta \Delta\Psi N)$ and $N_{m}|_{\Omega = 0}/(\Delta\Theta \Delta\Psi N)$, respectively. Here, $N_{m}$ is the number of pairs (among total $N$ pairs) satisfying the conditions $\theta_{n} \in [m \Delta\Theta - \Delta\Theta, m \Delta\Theta]$ and $\psi_{n} \in [\pi/2 - \Delta\Psi/2, \pi/2 + \Delta\Psi/2)$, in which the parameters $m$, $\Delta\Theta$ and $\Delta\Psi$ are chosen to be $m = \overline{1,155}$ and $\Delta\Theta = \Delta\Psi = \pi/155$ (note also that $\sum_{m} N_{m} =N$).
\begin{figure}
    \centering
    \includegraphics[height=5cm,width=8.5cm]{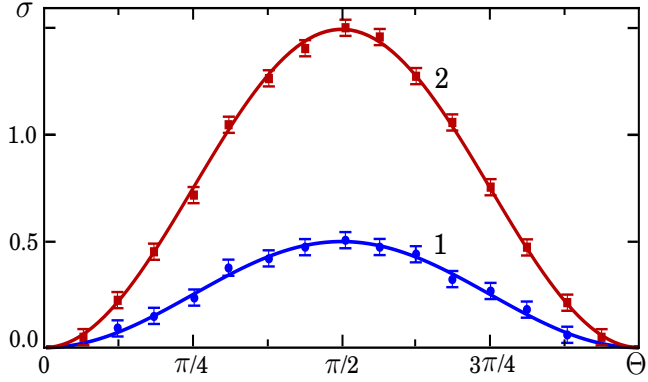}
    \caption{\label{fig4} (Color online) Dependence of
    $\sigma$ on $\Theta$ for $\Omega/h = 0.5$ (1) and
    $\Omega/h = 1.5$ (2). The solid curves correspond
    to the theoretical result $\sigma = (\Omega/h)
    \sin^{2}\Theta$ and the symbols represent the
    numerical values of $\sigma$. The latter are obtained
    by solving Eqs.~(\ref{eff_Lan2}) for $h = 2.5\times
    10^{-2}$, $\Omega = 1.25 \times 10^{-2}$ (1) and
    $\Omega = 3.75 \times 10^{-2}$ (2); the other
    parameters are given in the text.}
\end{figure}

As is illustrated in Fig.~\ref{fig4}, our simulation results for the $\Theta$-dependence of $\sigma$ are in a very good agreement with the theoretical prediction. This leads to the following conclusions. First, the solution (\ref{P_st2}) of the Fokker-Planck equation (\ref{FPst2}) correctly describes the long-time behavior of the rotational motion of nanoparticles in a viscous fluid. Second, since the difference $(P_{\mathrm{st}} - P_{0})|_{\Psi = \pi/2}$ is of the second order in $\kappa h$, the effective Langevin equations (\ref{eff_Lan2}) can be used to predict and study subtle rotational effects. And third, the representation $P_{\mathrm{st}} =P_{\mathrm{st}} (\Theta, \Psi)$, which is the key assumption in our analysis, holds not only at $\Omega/h < 1$ (as it could be expected from the noiseless case), but also at $\Omega/h \geq 1$. The last means that the periodic regime of rotation does not influence the steady-state probability density $P_{\mathrm{st}}$.

The numerical solution of Eqs.~(\ref{eff_Lan2}) is then used to determine the average values of the nanoparticle magnetization and nanoparticle angular frequency, $\langle \mu \rangle$ and $\langle \chi \rangle$. They are calculated as $\langle \mu \rangle = (1/l)\sum_{i=1}^{l} \cos \theta_{i}$ and $\langle \chi \rangle = 1 - (1/l) \sum_{i=1}^{l}\psi_{i} /(\Omega\tilde{t}_ \mathrm{sim})$, where $\theta_{i} = \theta_{i}(\tilde{t}_ \mathrm{sim})$ and $\psi_{i} = \psi_{i}( \tilde{t}_ \mathrm{sim})$ are the polar and lag angles in the $i$-th run, $\tilde{t}_ \mathrm{sim}$ is the simulation time, and $l$ is the total number of runs. In our simulations we set $\tilde{t}_ \mathrm{sim} = 10^{6}$ and $l = 10^{5}$; the other parameters are the same as in Fig.~\ref{fig2}. It should be noticed that since $\tilde{t}_ \mathrm{sim}$ is large enough, the statistical properties of angles $\theta_{i}$ and $\psi_{i}$ do not depend on their initial values.

Using this approach, we observed that $\langle \mu \rangle =0$ for all finite values of the inverse temperature parameter $\kappa$. At first sight, this result is in disagreement with the behavior of $\mu$ in the noiseless case (when $\kappa = \infty$), see Fig.~\ref{fig2}, because $\langle \mu \rangle$ at large $\kappa$ should approach $\mu$. If $\mu$ is numerically determined for small but non-zero values of $h_{z}$ (e.g., $h_{z} = 10^{-3}$ in Fig.~\ref{fig2}), then $\langle \mu \rangle$ at $\kappa \gg 1/h_{z}$ indeed approaches $\mu$. However, since $\mu$ is mathematically defined as $h_{z} \to 0$, such an approach is impossible for any finite $\kappa$. In fact, the periodic regime of nanoparticle rotation, which exists in the noiseless limit at $h_{z}=0$ and $\Omega/h\geq 1$, is degenerate: $\lim_{h_{z} \to 0^{\pm}} m_{z} = \pm \mu$. The thermal torque of arbitrary strength completely destroys this regime, leading to $\langle \mu \rangle =0$.

\begin{figure}
    \centering
    \includegraphics[height=5cm,width=8.5cm]{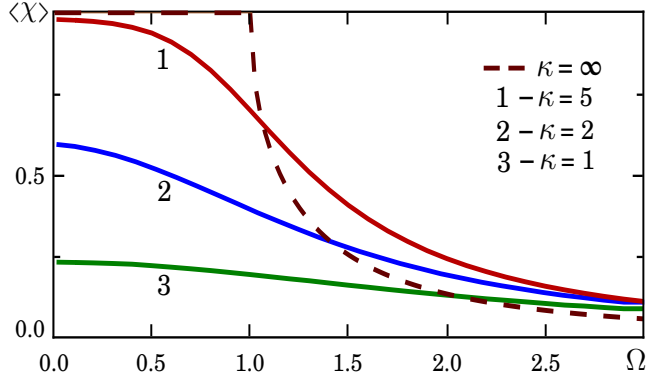}
    \caption{\label{fig5} (Color online) Average value
    of the reduced angular frequency of nanoparticles,
    $\langle\chi \rangle$, as a function of $\Omega$
    for $h_{z}=0$, $h=1$ and different values of the
    parameter $\kappa$. The solid curves show the results
    obtained by numerical solution of Eqs.~(\ref{eff_Lan2}),
    and the dashed curve represents Eq.~(\ref{chi}), which
    corresponds to the noiseless case ($\langle \chi \rangle
    |_{\kappa = \infty} = \chi$).}
\end{figure}

\begin{figure}
    \centering
    \includegraphics[height=5cm,width=8.5cm]{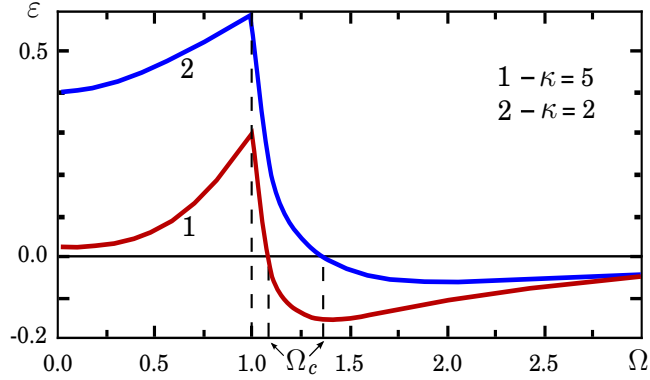}
    \caption{\label{fig6} (Color online) Dependence of
    $\varepsilon$ on $\Omega$ for different values of
    $\kappa$. The simulation parameters are chosen
    as in Fig.~\ref{fig5}.}
\end{figure}

The influence of thermal torque on the average angular frequency of rotation of nanoparticles driven by a circularly polarized magnetic field (when $h_{z}=0$) is illustrated in Fig.~\ref{fig5}. As seen, the average angular frequency $\langle \chi \rangle$ is strongly affected by thermal torque (the less the parameter
$\kappa$, the more the torque strength) and exhibits a remarkable dependence on the driving field frequency $\Omega$. Since $\langle \tilde{\mu}_{1,2} (\tilde{t}) \rangle = 0$, see Eqs.~(\ref{eff_Lan2}), the dependence of $\langle \chi \rangle$ on $\kappa$ and $\Omega$ is a purely nonlinear effect. Its most striking manifestation is that thermal fluctuations can both increase and decrease the angular frequency of nanoparticles as compared with the deterministic case. Specifically, if $\Omega_{c}$ is the solution of the equation $\varepsilon = 0$ ($\varepsilon  = \langle \chi \rangle|_{\kappa = \infty} - \langle \chi \rangle$) with respect to $\Omega$, then thermal fluctuations decrease the frequency of rotation ($\varepsilon >0$) when $\Omega < \Omega_{c}$, and increase it ($\varepsilon < 0$) when $\Omega > \Omega_{c}$ (see Fig.~\ref{fig6}). Note that $\Omega_{c}$ grows and $|\! \min \varepsilon|$ decreases as $\kappa$ becomes smaller, and $\varepsilon$ approaches zero at large $\Omega$.

By solving the effective Langevin equations (\ref{eff_Lan2}) numerically, we investigate the role of thermal torque in the nanoparticle dynamics induced by the precessing magnetic field (when $h_{z} \neq 0$). Before we proceed with the analysis of thermal effects, we recall that in the noiseless case the steady-state dynamics of nanoparticles has a precessional character described by constant polar and lag angles (\ref{psi_theta}). As a consequence, in this case $\chi=1$, i.e., the angular frequency of precessional rotation of nanoparticles coincides with the magnetic field frequency, and the $z$-component of the reduced nanoparticle magnetization is given by Eq.~(\ref{m_z}).

\begin{figure}
    \centering
    \includegraphics[height=5cm,width=8.5cm]{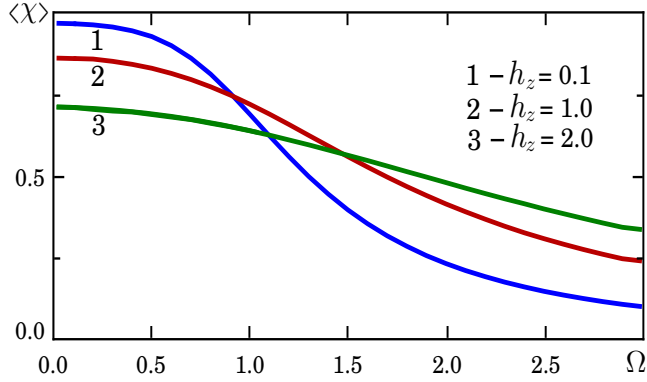}
    \caption{\label{fig7} (Color online) Average value
    of the reduced angular frequency of precession of
    nanoparticles as a function of $\Omega$ for $h=1$,
    $\kappa = 5$ and different values of the parameter
    $h_{z}$. }
\end{figure}

\begin{figure}
    \centering
    \includegraphics[height=5cm,width=8.5cm]{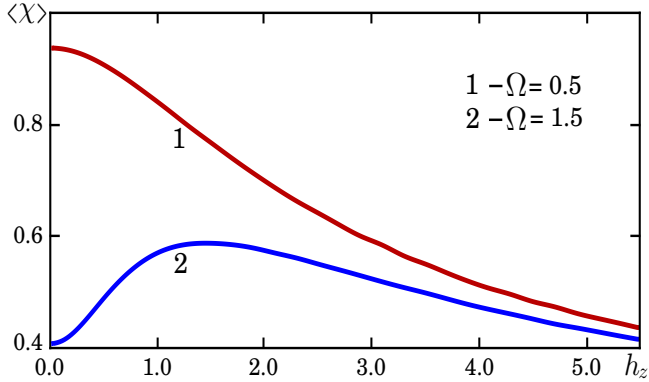}
    \caption{\label{fig8} (Color online) Average value
    of the reduced angular frequency of precession of
    nanoparticles as a function of $h_{z}$ for $h =1$,
    $\kappa = 5$ and two values of $\Omega$. In this
    case, the characteristic frequency $\Omega_{0}$
    that separates the regimes of monotonic ($\Omega <
    \Omega_{0}$) and non-monotonic ($\Omega >\Omega_{0}$)
    dependence of $\langle \chi \rangle$ on $h_{z}$ is
    equal to $0.78$.}
\end{figure}

Because Eqs.~(\ref{eff_Lan2}) are nonlinear, the thermal torque essentially influences the average characteristics of the precessional motion of nanoparticles. In particular, due to its action, the average angular frequency of precession becomes less than the magnetic field frequency, i.e., $\langle \chi \rangle < 1$ for all finite values of $\kappa$. Moreover, the numerical simulations show that $\langle \chi \rangle$ is a monotonically decreasing function of $1/\kappa$ with $\langle \chi \rangle| _{1/\kappa =0} = 1$ and $\langle \chi \rangle| _{1/\kappa =\infty} = 0$. The average frequency $\langle \chi \rangle$ also decreases monotonically with increasing $\Omega$ (see Fig.~\ref{fig7}), and $\langle \chi \rangle \to 0$ as $\Omega \to \infty$ for each finite $\kappa$. An important
feature of these dependencies is that they decrease more slowly with increasing $h_{z}$. This fact suggests the existence of the characteristic frequency $\Omega_{0}$, which separates two qualitatively different behaviors of $\langle \chi \rangle$ as a function of $h_{z}$. Namely, if $\Omega < \Omega_{0}$ then $\langle \chi \rangle$ monotonically decreases as $h_{z}$ increases, and $\langle \chi \rangle$ exhibits a non-monotonic dependence on $h_{z}$ if $\Omega > \Omega_{0}$, as shown in Fig.~\ref{fig8}. It is important to emphasise that all
these remarkable properties of the average frequency of precession of nanoparticles result from thermal fluctuations; in the noiseless case $\langle \chi \rangle = 1$.

\begin{figure}
    \centering
    \includegraphics[height=5cm,width=8.5cm]{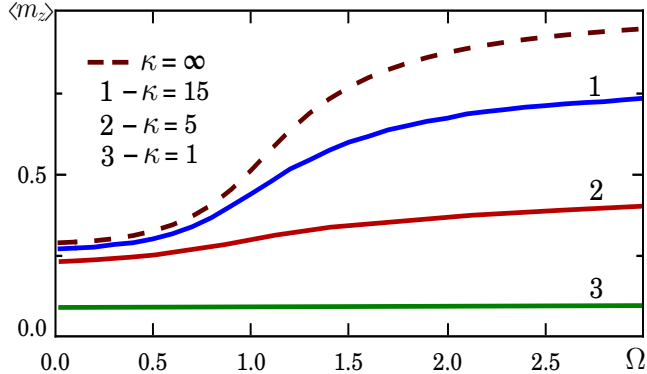}
    \caption{\label{fig9} (Color online) Frequency
    dependence of the average reduced magnetization
    for $h_{z} = 0.3$, $h=1$ and different values of
    the parameter $\kappa$. The dashed curve represents
    the theoretical result (\ref{m_z}) for the noiseless
    case.}
\end{figure}

Finally, the dependence of the average reduced magnetization $\langle m_{z} \rangle$ on $\Omega$ and $\kappa$ is illustrated in Fig.~\ref{fig9}. As seen, $\langle m_{z} \rangle$ approaches the theoretical result (\ref{m_z}) as $\kappa$ grows, and $\langle m_{z} \rangle$ almost does not depend on $\Omega$ at relatively small $\kappa$. Since the limit $\Omega \to \infty$ corresponds to the absence of the rotating magnetic field, from Eq.~(\ref{P_st}) one obtains $\langle m_{z} \rangle|_{\Omega \to \infty} = 2\pi \int_{0}^{\pi} \cos \theta P_{0}(\theta) d\theta = L(\kappa h_{z})$, where $L(x) = \coth x - 1/x$ is the Langevin function. In particular, for curves 1, 2 and 3 the function $L(\kappa h_{z})$ approximately equals $0.778$, $0.438$ and $0.099$, respectively.

\section{CONCLUSIONS}
\label{Concl}

We have studied both analytically and numerically the rotational properties of ferromagnetic nanoparticle in a viscous fluid driven by a precessing magnetic field. Our approach is based on the system of multiplicative Langevin equations for the polar and azimuthal angles of the nanoparticle magnetization frozen into the massless nanoparticle. From these equations, approximating the Cartesian components of the random torque by Gaussian white noises and interpreting them in an arbitrary way, we have derived the corresponding Fokker-Planck equation. By associating the stationary solution of this equation with the Boltzmann probability density, we have established that the basic system of Langevin equations should be interpreted in the Stratonovich sense. Within this framework, we have reproduced the known system of effective Langevin equations, which is simpler than the basic one, and have shown that the statistical properties of its solution do not depend on the interpretation of multiplicative white noises.

Using the system of effective Langevin equations and the corresponding Fokker-Planck equation, we have calculated the average angular frequency of precession of nanoparticles and the average magnetization of nanoparticles in the $z$-direction, and have analyzed their dependence on the model parameters. In the noiseless limit, the dependence of these quantities on the rotating field frequency and amplitude is different whether a constant component of the magnetic field is zero or not. In the former case, it has been shown both analytically and numerically that the angular frequency and magnetization depend only on the ratio of the rotating field frequency to the rotating field amplitude and, starting from a certain value of this ratio, these dependencies become strongly nonlinear. The most remarkable property of the above mentioned quantities is that their sum is strictly equal to 1 (in dimensionless units) for any rotating field. In the latter case, when the steady-state rotation of nanoparticles has the precessional character, we have derived a general expression for the nanoparticle magnetization and have observed that the frequency of nanoparticle precession always coincides with the rotating field frequency.

The influence of thermal fluctuations on the rotational dynamics of nanoparticles is investigated by numerical integration of the system of effective Langevin equations. To verify these equations, we first calculated the difference between the steady-state and equilibrium probability densities of the nanoparticle orientation, which arises from a slowly rotating magnetic field of small amplitude. Then, by comparing the numerical results for this difference with the results obtained from the analytical solution of the Fokker-Planck equation, we have confirmed the validity of effective Langevin equations. Finally, using these equations, we have observed a number of interesting thermal effects. In particular,  it has turned out that the deterministic regime of rotation of nanoparticles, which exists when a constant magnetic field is infinitesimally small and the rotating field frequency exceeds the critical one, is completely destroyed by thermal fluctuations. But the most important observation is that thermal fluctuations can play a constructive role in the precessional dynamics of nanoparticles. The non-monotonic behavior of the average angular frequency of nanoparticle rotation as a function of the constant magnetic field strength supports this statement.

\section*{ACKNOWLEDGMENTS}

V.V.R. acknowledges the Erasmus Mundus programme for financial support.

\end{document}